# The NewAthena mission concept in the context of the next decade of X-ray astronomy

A list of authors and their affiliations appears at the end of the paper

Large X-ray observatories such as Chandra and XMM-Newton have been delivering scientific breakthroughs in research fields as diverse as our Solar System, the astrophysics of stars, stellar explosions and compact objects, accreting supermassive black holes, and large-scale structures traced by the hot plasma permeating and surrounding galaxy groups and clusters. The recently launched X-Ray Imaging and Spectroscopy Mission observatory is opening in earnest the new observational window of non-dispersive high-resolution spectroscopy. However, several questions remain open, such as the effect of the stellar radiation field on the habitability of nearby planets, the equation of state regulating matter in neutron stars, the origin and distribution of metals in the Universe, the processes driving the cosmological evolution of the baryons locked in the gravitational potential of dark matter and the impact of supermassive black hole growth on galaxy evolution, to mention just a few. Furthermore, X-ray astronomy has a key part to play in multimessenger astrophysics. Addressing these questions experimentally requires an order-of-magnitude leap in sensitivity, spectroscopy and survey capabilities with respect to existing X-ray observatories. This article succinctly summarizes the main areas where high-energy astrophysics is expected to contribute to our understanding of the Universe in the next decade and describes a new mission concept under study by the European Space Agency, the scientific community worldwide and two international partners (JAXA and NASA), designed to enable transformational discoveries: NewAthena. This concept inherits its basic payload design from a previous study carried out until 2022, Athena.

Cosmological simulations suggest that most of the baryons in the universe are 'hot'[1], that is, in a temperature and density regime where copious X-rays are emitted by thermal particles. Consequently, sensitive X-ray observations are required to ultimately answer several fundamental questions in modern astrophysics. These observations involve the study of systems as diverse as individual stars and their planetary environments, the results of stellar explosions (such as neutron stars, stellar-mass black holes and supernova remnants) and hot gas halos surrounding individual galaxies or permeating the space among galaxies in groups and clusters, eventually connecting to the cosmic web. Furthermore, supermassive black holes (SMBHs) most probably play a key role in shaping the cosmological evolution of their host galaxies, ultimately driving their rate of star formation. The quest for the root cause of this elusive 'feedback mechanism' requires measuring both the populations and the energetics of accreting SMBHs at the centre of galaxies, active galactic nuclei (AGNs). With this scope, X-ray measurements are indispensable.

✉e-mail: Matteo.Guainazzi@esa.int

In the 2030s, a suite of large multiwavelength astronomical facilities will be operational or will have surveyed the sky at unprecedented sensitivity from radio to very high-energy γ-rays. Furthermore, multimessenger astrophysics is expected to reach full maturity in the second half of the next decade, with the deployment of the third generation of ground-based gravitational wave arrays, new neutrino facilities[2,3] and the space-borne gravitational wave observatory LISA[4]. Explosive and transient phenomena in the Universe are often associated with the emission of high-energy radiation. Sensitive X-ray observations of neutrino- and gravitational-wave-emitting sources are a key tool of multimessenger astrophysics.

An X-ray observatory matching and complementing this suite of facilities will therefore allow us to uniquely address a set of fundamental questions in modern astrophysics, such as the following.

- How does the stellar radiation field affect the habitability of planetary systems, and how is it in turn influenced by the presence of nearby planets?
- What is the equation of state regulating matter in neutron stars?
- What is the origin of the high-energy processes in the close environment of black holes?
- What distribution of supernovae and supernova explosions leads to the mixture of metals we measure in the local Universe? How are metals distributed through the cosmos?
- What drives the cosmological co-evolution of galaxies and SMBHs?
- How does SMBH feedback shape the large-scale baryon distribution?
- How do large-scale structures in the Universe form and evolve? What physics defines their hot gas content?
- What is the astrophysical nature of the most common celestial sources of neutrinos and gravitational waves?

These questions remain open despite the enormous advances brought by past and operational flagship X-ray observatories such as Chandra and XMM-Newton[5]. In this Perspective, we explore the enhancement in science performance that will allow these open questions to be addressed. In the last section, we advocate a new mission concept under study by the European Space Agency (ESA) and in the science community worldwide, capable of achieving this science performance: NewAthena.

## How do black holes grow and influence galaxy evolution?

It is now well established[6] that most massive galaxies host at their centres an SMBH, with mass of ≥$10^6 M_\odot$. The SMBH masses are well correlated with the masses of their host galaxies (with a better correlation with their central parts)[7]. This leads to the question of how a system with the size of the Inner Solar System (the typical scale of the event horizon of the central SMBH) could affect (or perhaps even control) phenomena on scales millions to billions of times larger. This mystery is compounded by the parallel evolution across cosmic time of the growth of galaxies via star formation and the growth of SMBHs via accretion of material from their surroundings, a process enabling them to shine as AGNs[8]. Both processes were much stronger in the past, with a broad peak around redshift $z \approx 1-3$ (so-called Cosmic Noon), and a fast decrease towards the present date. The exact behaviour at higher $z$ is strongly debated: galaxy growth seems to have had a slower increase sustained in time from high $z \approx 10$, while AGN power may have grown faster starting at lower $z$, although the recent detection by the James Webb Space Telescope of luminous AGNs with SMBHs of masses $\sim 10^7-10^9 M_\odot$ at $z > 5$ challenges current models[9]. Whatever the explanations for the facts above, the growth of galaxies and of the SMBHs in their centres must be inextricably connected, in what is called the co-evolution of AGNs and galaxies[10], but the physical processes that drive such co-evolution remain poorly understood. The stupendous amount of radiation from the AGNs can generate energetic outflows of material[11], perhaps pushing or heating the interstellar and intergalactic medium to the point where star formation is no longer possible or sparking star formation[12]. Alternatively (or perhaps complementarily), galactic star formation and SMBH growth could be controlled by the flow of intergalactic gas into galaxies[13], possibly enhanced by mergers and/or by external fuelling[14]. Understanding the physical and astrophysical drivers of the AGN–galaxy co-evolution is one of the main topics of current extragalactic astronomy.

In accreting black hole systems over the whole mass range, an X-ray telescope with a collecting area substantially larger than currently operational observatories will enable studies of the highly variable inner accretion flow close to its dynamical timescale in AGNs (~hours). These studies will, for example, allow us to establish the location and nature of the primary source of hard X-rays, and its connection to the typically observed fast outflows, expected to be launched from the inner accretion disk. Such observations will also measure the black hole spin to a high precision. The distribution of spin in local Universe AGNs is a sensitive probe of their growth processes[15].

Strong energetic relativistic outflows from the inner region around SMBHs have been measured in X-rays[11]. It is crucial to understand how such outflows around SMBHs are launched, and how they are connected to the larger-scale feedback in the surrounding galaxies. Sensitive, time-resolved X-ray spectroscopic observations are indispensable to determine the outflow energy and mass rates from the local Universe to Cosmic Noon, fundamental quantities that remain impossible to constrain over a sizeable fraction of the AGN population, even in the deepest AGN observations with the Chandra and XMM-Newton spectrometers. Furthermore, the impact of such AGN-driven outflows up to Cosmic Noon can be gauged by identifying a spectral signatures in the overall AGN population through a sufficiently deep extragalactic survey. This is beyond the capabilities of even the most powerful X-ray survey mission flown to date, eROSITA[16] (cf. the right panel of Fig. [1]).

The space density of X-ray-detected AGNs exceeds that of radio-, UV- or optically selected AGNs by a large factor[17–19]. Furthermore, X-rays can pierce the obscuration known to exist around the centres of many of these objects[20]. A large-area sensitive X-ray extragalactic survey would constrain the overall SMBH accretion rate density, reaching AGNs around the knee of the luminosity function (where most of the black hole mass growth occurs), out to the epoch of reionization for unobscured objects, and characterizing moderate to intermediate obscuration up to $z \approx 6-7$ (which may dominate accretion growth[21]; cf. the left and central panels of Fig. [1]). This critical part of the parameter space is difficult to reach for facilities at other wavelengths with sufficient statistics. Other spectral ranges and facilities (for example, the Square Kilometre Array) would be able to detect high-$z$ heavily obscured AGNs in great numbers[22,23], but their characterization and recognition as such would require extensive additional multiwavelength data (see, for example, synergies with future European Southern Observatory missions as described in ref. [24]). This leads back to the key role that an X-ray observatory capable of providing a wide census of the AGN population through cosmic ages would have in this quest.

## X-ray emission from neutron stars as a probe of dense-matter physics and multimessenger astrophysics

The detection of gravitational waves emanating from the binary neutron star merger GW170817, followed by multiband observations of its electromagnetic counterparts, where X-rays played a crucial role[25], marked the advent of multimessenger astronomy. In the forthcoming decades, we anticipate similar events occurring at a frequency of almost one per day when the next generation of gravitational wave interferometers becomes operational[26]. X-ray measurements of jets and their angular structure and orientation can disentangle the distance and inclination of a gravitational wave source, thereby enhancing the precision of cosmological inferences. Furthermore, X-ray observations can track the activity of the merger remnant and the emergence of kilonova afterglows[27], providing novel constraints on the behaviour of dense matter before and after the merger. This requires a large-area X-ray facility matching the technological development of ground-based gravitational wave arrays[28].

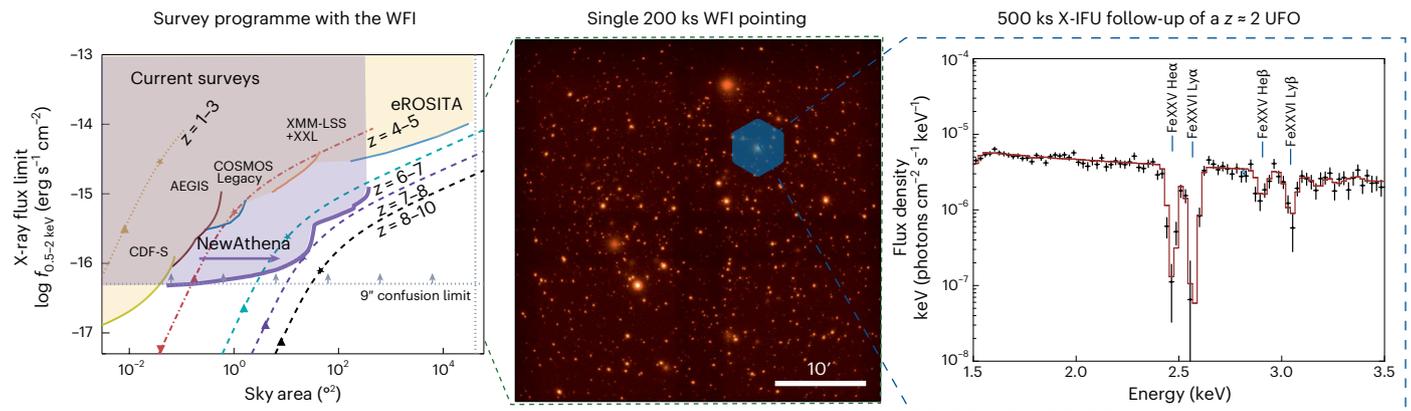

**Fig. 1 | Illustration of the survey capability of the NewAthena WFI.** Left: X-ray limiting flux versus sky coverage for a representative sample of current X-ray surveys (yellow shaded area), compared with the improvement enabled by the NewAthena WFI (purple shaded area). The dashed lines indicate the area coverage to different flux limits required to reveal statistical samples of AGNs in various redshift ranges, with stars indicating the $L^*$ where most growth occurs. The grey dotted horizontal line indicates the flux corresponding to the confusion limit for the NewAthena telescope on-axis point spread function (9″ HEW). Centre: simulation of a 200 ks WFI observation of an extragalactic field with Simulation of X-ray Telescopes (SIXTE)[55]. It is based on the Chandra Deep Field South. The blue hexagon represents the field of view of the X-IFU instrument, targeting a candidate AGN with an ultrafast outflow identified from the WFI survey. The spectrum indicates a 500 ks X-IFU simulation of an ultrafast outflow in an AGN at $z \approx 2$ around the transitions of He- and H-like iron. The outflow parameters are column density $N_H = 10^{24}$ cm$^{-2}$, ionization parameter log($\xi$) = 4; velocity $v/c = 0.1$. The error bars in the spectrum are at the 1$\sigma$ level.

Likewise, with a large X-ray observatory, we may finally approach the longstanding goal of constraining the equation of state for dense matter. This is a fundamental challenge in both physics and astrophysics. One promising approach has long been recognized: measuring the mass and radius of a neutron star. While most neutron stars are identified as radio pulsars, a select few exhibit periodic X-ray modulations of emitted radiation from their surfaces, a direct consequence of their rotation and their high surface temperature. If this periodic emission stems from one (or multiple) hot spot(s) on the star's rotating surface, we can predict the emission observed by a distant observer using a technique known as ray-tracing. This method traces the path of light from the neutron star's surface to the observer through the curved spacetime around the star, with the effects of general relativity encoded in the resulting periodic emission observed from the neutron star's surface[29]. The light-curve model derived from this approach can then be compared with observations to constrain various parameters, including the neutron star's mass and radius. The Neutron Star Interior Composition Explorer mission has pioneered this kind of measurement[30,31]. However, the accuracy in the determination of the neutron star radius (10–15%) is still insufficient to constrain the equation of state. Only a large X-ray observatory with a low and well-characterized internal background will be able to measure the mass and radius of a large sample of neutron stars at the percentage level required to provide a strong constraint on the dense-matter equation of state.

## Mapping the dynamical assembly of intergalactic plasma in the large-scale structure

The structure we observe today on the largest scales of the cosmos originates from tiny density perturbations left after the Big Bang. Under the influence of gravity, small overdense clumps of matter have merged over time, leading to a web-like structure of galaxies, galaxy groups, galaxy clusters and large-scale filaments, spanning the observable Universe[32]. This succession of mergers injects kinetic energy into the newly formed structures, which is eventually dissipated into heat. This is why most of the normal, baryonic matter is today in the form of a diffuse plasma with temperatures reaching millions to hundreds of millions of degrees, filling the space around and between galaxies in the cosmic web. X-ray astronomy offers one of the most detailed physical diagnostics of this hot intergalactic medium/intracluster medium (ICM) through its emission and absorption signatures.

Whether in the form of shock fronts or turbulence driven during the ongoing mergers, it is expected that the signatures of the structure-formation process are most directly traced by the gas velocity field. This can be directly probed, in principle, by imaging the Doppler shifts and broadening of X-ray emission lines from the ICM. However, such measurements have remained largely out of reach for previous X-ray observatories. Therefore, little is known about how the kinetic energy is injected and eventually thermalized during large-scale structure growth. Recently, the X-Ray Imaging and Spectroscopy Mission (XRISM) has inaugurated the era of electronvolt-level, non-dispersive X-ray spectral imaging over the 1.7–12 keV energy band[33] with its microcalorimeter instrument Resolve[34]. While this will give us valuable insight into the gas dynamics in local, bright, hot clusters of galaxies, an additional leap in effective area and spatial resolution is needed to effectively map bulk motions and turbulence over a substantial fraction of these objects' volume, and to trace the evolution of the ICM dynamics to higher redshift. An improved sensitivity to low-surface-brightness emission would further allow us to probe the gas entropy at the outer edges of galaxy clusters, reaching much higher redshifts than current facilities, obtaining a complementary test of how the heating processes evolve over cosmic time.

Shocks and turbulence generated during galaxy cluster mergers are also believed to lead to the acceleration of a small fraction of particles to relativistic energies[35]. Mapping the corresponding gas motions with a powerful X-ray spectro-imaging instrument thus offers a key to understanding the creation of cosmic-ray electrons and the amplification of magnetic fields, detected in the radio band in the form of radio halos and radio relics.

Once heated by the gravitational assembly process, most of the gas would take longer than the Hubble time to radiatively cool and condense into stars. A notable exception is in the dense, bright centres of cool-core galaxy clusters. However, despite the short cooling times, it was observed that the star formation here remains inefficient: it is widely believed that AGN–ICM interaction provides the energy needed to prevent the gas from cooling. Understanding this dynamical process requires us to probe the associated gas kinematics. None of the existing operational X-ray missions can achieve this measurement at the required level of spatial and spectral resolution, even in very close objects such as M87. This field requires non-dispersive spectroscopy as well as an order-of-magnitude advancement in effective area and spatial resolution with respect to Resolve[34], whose optics exhibit only a moderate spatial resolution (~1′ half energy width, HEW); cf. Fig. 2 for an illustration of the required performance enhancement.

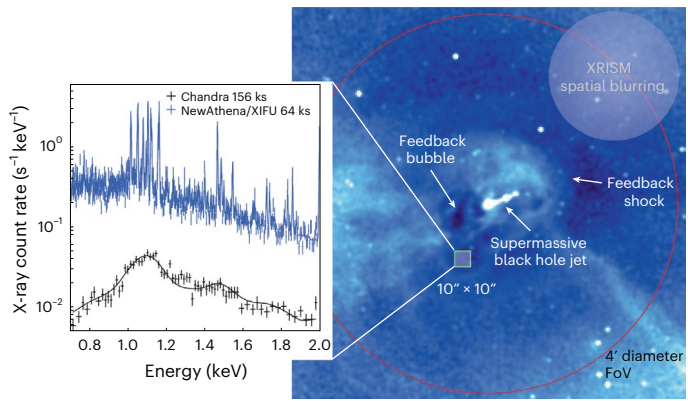

**Fig. 2 | NewAthena X-IFU spectrum of a feedback bubble in M87.** Left: the X-IFU simulated spectrum compared with an archival Chandra/ACIS observation of the same region. The spectra are extracted from the grey square in the ACIS image. The red circle represents the X-IFU field of view. The light-blue circle represents the point spread function of XRISM/Resolve, whose size is comparable to the region mostly affected by the interaction between the relativistic jet and the ICM. The error bars in the spectrum are at the 1σ level. FoV, field of view.

Far from being limited to the brightest cluster galaxies, feedback from AGNs and supernovae is now recognized as a lynchpin in the evolution of all galaxies. At the characteristic luminosity, $L*$, mass regime, corresponding to more 'typical' galaxies, much less is known about the feedback substructures, because these gaseous halos are substantially fainter in X-rays when compared with the ICM. The eROSITA 'bubbles' above and below the Galactic plane have been interpreted as evidence of past episodes of energetic feedback[36], and some of our state-of-the-art models of galaxy formation predict that these features are common in Milky Way/M31 analogues at $z = 0$ (ref. [37]). An observatory carrying X-ray optics with an HEW of ~10″ and an effective area about one order of magnitude larger than those of Chandra and XMM-Newton would enable us to search for such feedback bubbles in other nearby galaxies, therefore testing this model prediction and informing the future development of galaxy evolution codes.

Feedback processes result in heating the gas and, especially in lower-mass systems, also expelling it to large distances from the host galaxy, sometimes unbinding it completely. This leaves an imprint in how the gas mass fraction varies with halo mass; for large samples of low-mass systems out to large redshifts, this effect is most easily probed indirectly by, for example, looking at the luminosity–temperature relation. A survey conducted by a sensitive, wide-field X-ray imager would enable the detection of several thousand galaxy groups ($M_{500} < 5 \times 10^{13} M_\odot$) at $z \geq 1$ with accurate measurements of temperature and luminosity. Constraints in this mass and redshift regime, which is beyond the capability of future surveys based on the Sunyaev–Zeldovich effect (for example, CMB-S4) due to their modest angular resolution, would provide critical information for understanding and modelling galaxy, galaxy group and cluster evolution, and for accurately applying cosmological tests that require understanding of the nonlinear regime—for instance, those based on gravitational lensing.

How is the gas ejected by feedback eventually distributed? Where does it end up and how does it circulate through the veins and tendrils of the cosmic web? To answer this question, we ultimately need to detect the most diffuse hot plasma located in the outskirts of $L*$ galaxy halos and extending beyond the bounds of virialized structures into the large-scale structure filaments that connect them. These are the most common baryonic reservoirs in the Universe, wherein most of the normal matter resides. A sensitive, wide-field-of-view soft-X-ray imaging telescope will enable detection of several instances of the brightest and hottest large-scale structure filaments connecting to massive local galaxy clusters. Recent eROSITA results[38] demonstrate that such systems exist. Going to even lower gas densities, such as those expected in the outskirts of the circumgalactic medium of typical $L*$ galaxies, the surface brightness is predicted to drop far below that of the Milky Way foreground. This emission is predominantly in the form of spectral lines[39], in particular the He-like Kα oxygen transitions at a rest-frame energy of 0.57 keV. A detector with a spectral resolution of a few electronvolts would enable all lines from the O VII triplet for a target at $z \geq 0.035$ to be cleanly separated from the corresponding Milky Way line transitions, enabling the detection of much lower-density gas than is possible in broadband imaging. Combined with a field of view of a few arcminutes, such capabilities would open a new window on the properties of the circumgalactic medium on scales of ~100 kpc from the galaxy centres. This provides arguably the cleanest test of existing galaxy formation models, because the modelling uncertainties related to treating the (even more complex) central interstellar medium are minimized.

Finally, the most diffuse gas in the large-scale structure can be efficiently studied through the absorption features against bright background sources. The search for this gas component requires a combination of effective area and energy resolution far exceeding the capability of existing X-ray spectrometers, despite a claim of marginal warm–hot intergalactic medium detection in extremely deep observation with the XMM-Newton Reflection Grating Spectrometers (RGS)[40]. Such measurements would allow us to determine where most of the normal matter resides in the local Universe, providing a solution to the decade-long quest for the 'missing baryons'.

## Probing the evolution of metal factories in the Universe

The chemical enrichment history of the Universe is a broad topic with multidisciplinary appeal. The ICM represents the integrated enrichment averaged over billions of supernova explosions, making it a particularly clean probe to test cosmic nucleosynthesis. We need an important improvement in instrument capability and control of the background (at a level of a few per cent) to map the Fe abundance measurements for local clusters routinely over their entire volume out to $R_{500}$, and down to the mass scale of groups of galaxies[41].

Detecting other chemical elements, in addition to Fe, is even more challenging. A full study of the evolution of metals in the ICM, constraining the relative yield of different classes of supernovae as well as details of their explosive mechanisms, requires high-resolution X-ray spectroscopic measurements coupled with a large effective area, enabling measurements of the chemical composition in galaxy clusters up to an epoch commensurate with Cosmic Noon. Abundances of rare elements such as potassium and titanium are of particular interest, because current supernova nucleosynthesis models heavily underpredict the abundances of these two elements, compared with Galactic archaeology constraints[42].

A complementary approach to study the origin of heavy elements in the Galaxy, as well as among the most powerful cosmic-ray accelerators, is to observe the explosion of stars via supernovae, and their remnants in the Galaxy. The non-thermal X-ray component of such remnants, which follows a power-law spectrum, gives important clues to the synchrotron emission due to the magnetic field and consequently the role of supernova remnants in generating Galactic cosmic rays. Only a large X-ray observatory will be able to study several such acceleration engines and to disentangle their geometry, element abundances and possible anisotropies related to the past explosion.

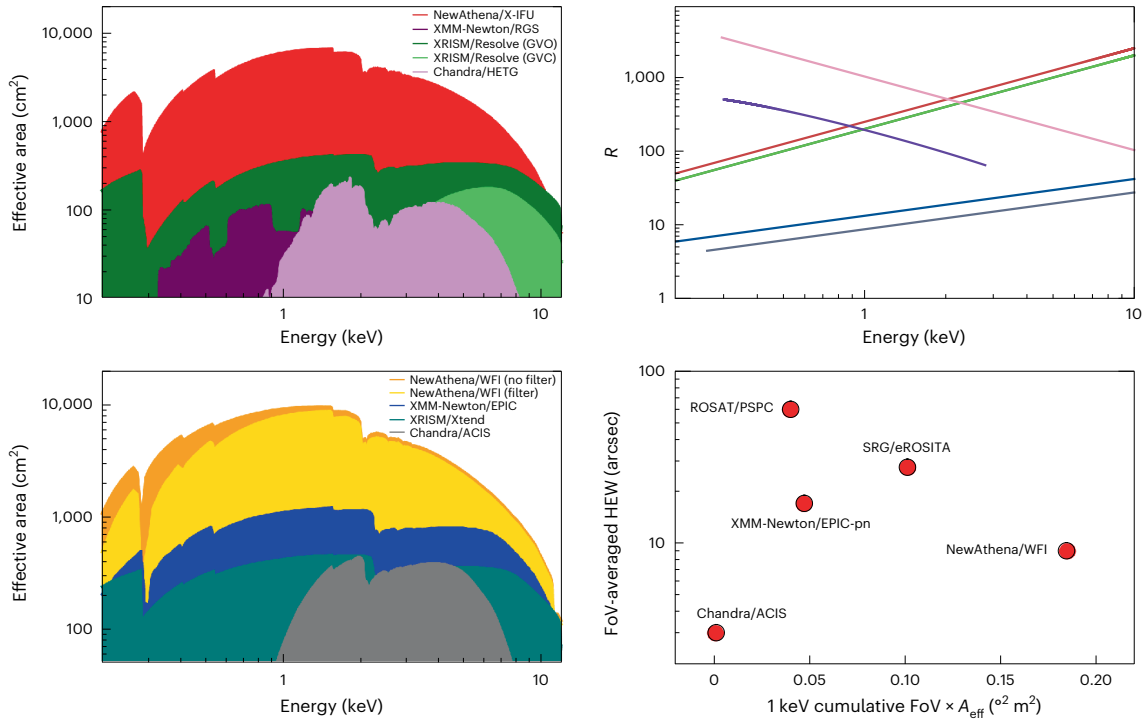

**Fig. 3 | Effective area of the NewAthena instruments and X-IFU resolving power $R$ as a function of energy, compared with operational high-resolution spectroscopic instruments.** $R = E/\Delta E$, where $\Delta E$ is the energy resolution. For XRISM/Resolve we show the configuration with the gate valve open (GVO; nominal requirement) and closed (GVC; current operational set-up). For RGS we show the area after the failure of two CCDs in 2000. The resolving power of two operational instruments based on CCDs (Chandra/ACIS and XMM-Newton/EPIC) is also shown for comparison. The colour scheme is the same in the resolving power panel as in the effective area panels—that is, in order of decreasing resolving power at 0.2 keV: Chandra/HETG, XMM-Newton/RGS, NewAthena/X-IFU, XRISM/Resolve, XMM-Newton/EPIC, Chandra/ACIS. Note that the resolving power of XRISM/Resolve is the same for the GVO and GCV configurations. The WFI grasp at 1 keV as a function of the field-of-view-averaged HEW is compared with past and operational survey X-ray instruments. Credit: bottom right, Arne Rau (MPE).

## Probing star–planet interaction with accurate X-ray spectroscopy

Almost 80% of stars may have hosted protoplanetary disks leading to the formation of planets. For the proximity of their habitable zones, the M-type stars are prime candidates for the characterization of potentially habitable exoplanets, and often show strong X-ray flares that may have an impact on the planet formation and their atmosphere. Giant planets appear to be scarce around M dwarfs[43], but terrestrial planets and super-Earths have an estimated occurrence rate approximately 3.5 times higher than around solar-mass stars[44,45]. This might be due to the star–planet interaction preferentially disrupting gas giant planets, something that X-ray emission can uniquely probe. A future large X-ray observatory might be able to diagnose flows in stellar coronae on minute timescales, spanning a factor of 20 in temperature, and can connect near-stellar activity with transient mass loss that would also impact planetary environments.

## NewAthena

ESA is leading the study of a mission concept able to address the scientific quests described in this Perspective. This concept, NewAthena, is a direct evolution of Athena, a mission selected in 2014 to address the scientific theme of the 'Hot and Energetic Universe'[46]. During Athena Phase A, a cohort of scientists and engineers at ESA, the Instrument Consortia, the International Partners (JAXA and NASA) and the broad science community have contributed to developing the science case for the Athena observatory. This Perspective heavily relies on the scientific case developed for Athena. We refer readers interested in the original Athena science cases to the white papers published together with ref. 46, as well as to refs. 47,48. NewAthena may join a fleet of new X-ray observatories possibly operational in the next decade such as

**Table 1 | NewAthena key scientific requirements**

| Parameter | Required value |
|---|---|
| X-IFU total effective area at 7 keV | $0.087\,m^2$ |
| X-IFU total effective area at 1 keV | $0.60\,m^2$ |
| X-IFU energy resolution at 7 keV | 4 eV |
| X-IFU field of view (effective diameter) | 4 arcmin |
| X-IFU pixel size on the sky | 5 arcsec |
| X-IFU background (2–7 keV) | $5\times10^{-3}\,photons\,cm^{-2}\,s^{-1}\,keV^{-1}$ |
| WFI effective area at 1 keV | $0.86\,m^2$ |
| WFI field of view (side) | 40 arcmin × 40 arcmin |
| WFI background (2–10 keV) | $8\times10^{-3}\,photons\,cm^{-2}\,s^{-1}\,keV^{-1}$ |
| Background knowledge accuracy | 5% |
| Optics angular resolution on axis at 1 keV | 9 arcsec |
| Field-of-view-averaged optics angular resolution at 1 keV | On axis +1 arcsec |
| Point source (45° off-axis) X-ray stray light area ratio against on-axis area | $1\times10^{-3}$ |
| Field of regard | 34% |
| Target of opportunity response time | 12 h |

These scientific requirements for NewAthena were endorsed by the Science Programme Committee of the ESA in November 2023. For comparison, the scientific requirements of the Athena mission are described in ref. 46, Table 4.

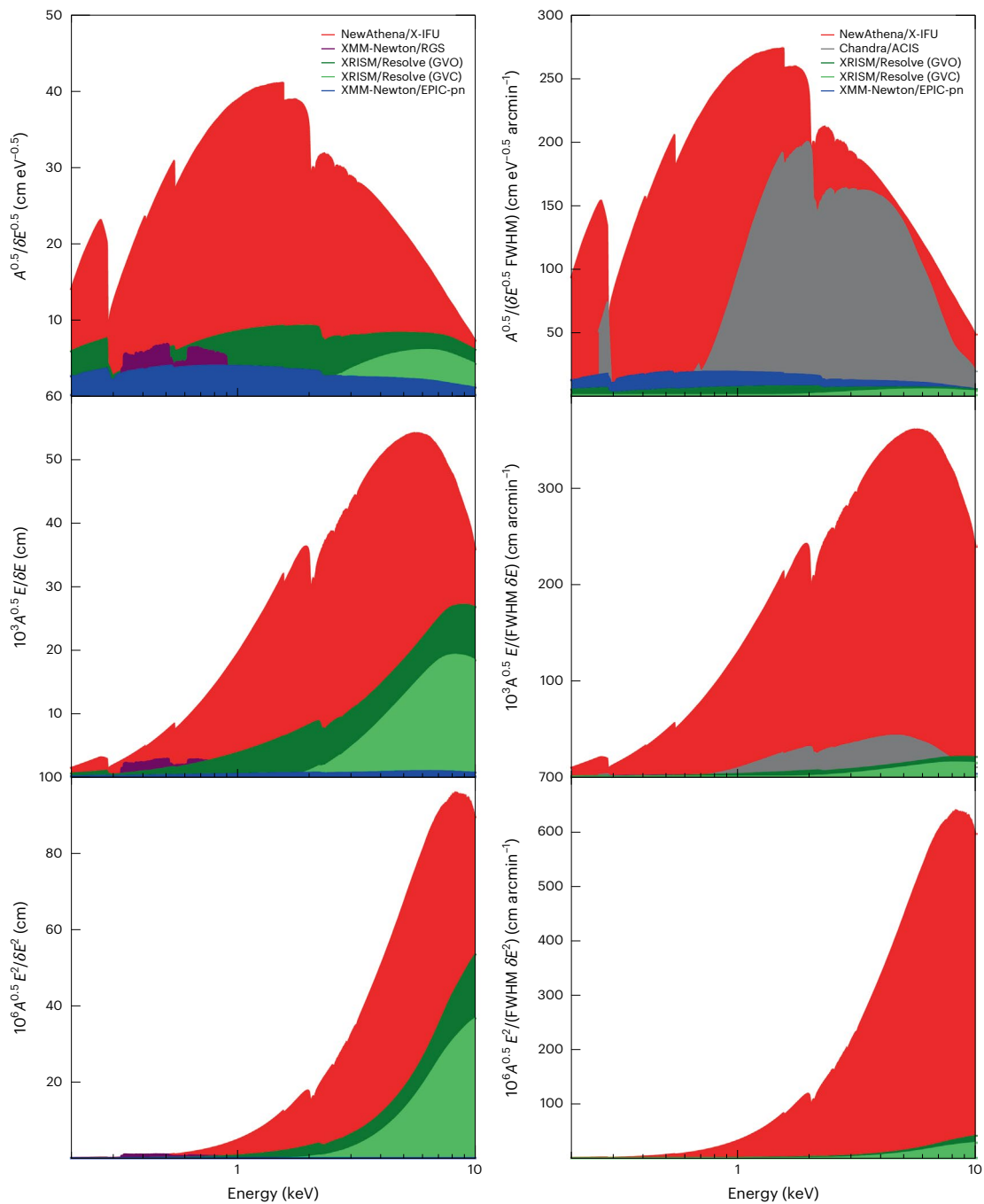

**Fig. 4 | Spectroscopic FoMs comparing the NewAthena X-IFU with operational X-ray spectrometers.** From top to bottom, detection of weak lines, and shift (velocity) and width (broadening) of strong lines. Left column: point-like sources. Right column: extended sources. The definitions are in the y-axis label. The extended-source FoMs are calculated from the point-like-source ones by dividing by a further factor equal to the telescope full-width at half-maximum to account for the sum in quadrature of the signal-to-noise ratio in a number of independent spatially resolved extraction regions.

the X-ray probe AXIS[49], or dedicated high-resolution spectroscopic facilities such as HUBS[50].

In 2022, ESA started a 'reformulation' of the mission profile and science case, because the estimated costs exceeded the level of resources available in the ESA Science Programme. The new concept carries the same scientific payload as on Athena: a Wide Field Imager (WFI)[51] based on an active silicon detector, with a 40′ × 40′ field of view and moderate (CCD (charge-coupled device)-like) energy resolution; and a pixelated X-Ray Integral Field Unit (X-IFU)[52], with unprecedented energy resolution ($\Delta E \leq 4$ eV) over more than 1,500 pixels, of about 5″ side each. The two instruments can be moved to the focal plane of a single-tube, 12-m-focal-length, telescope with a large effective area and an average ~10″ HEW angular resolution over the whole WFI field of view (~9″ on axis). The main scientific requirements of NewAthena are listed in Table 1.

The NewAthena telescope is based on silicon pore optics technology[53], characterized by the largest ratio between area and mass for space-qualified X-ray optics to date[54]. They enable an effective area of the two focal plane instruments exceeding that of operational X-ray observatories by an order of magnitude or more at 1 keV. Coupled with the large field of view, the large effective area ensures that the WFI grasp—the product of these two quantities—exceeds by a large factor even that of X-ray missions specifically designed to perform X-ray surveys such as eROSITA[16] (Fig. 3). The resolving power of microcalorimeter detectors greatly exceeds that of grating detectors above 2 keV (and, obviously, the resolving power of CCD detectors at all energies).

It is useful to compare the spectroscopic performances of different instruments by using suitable combinations of basic instrument science performance parameters. These spectroscopic figures of merit (FoMs) are proportional to the signal-to-noise ratio to detect an absorption or emission line. We compare in Fig. 4 the FoM for the detection of a weak line, and the measurements of velocity centroid (shift) and width (broadening) of a strong line. For all these FoMs, the NewAthena X-IFU exceeds the performance of existing imaging, dispersive and non-dispersive spectrometers by more than an order of magnitude over most of the sensitive bandpass.

NewAthena is planned to be launched in 2037. It is therefore an X-ray observatory matching the suite of large-scale observational facilities operational in the 2030s, and providing the required combination of sensitivity, energy resolution and field of view that will enable transformational progress in all the scientific fields discussed in this Perspective, and undoubtedly many more.

## Data availability

The Chandra/HETG spectrum shown in Fig. 2 has been extracted from data available in public archives. All the NewAthena simulations shown in this Perspective were generated using public telescope and instrument responses made available by ESA and by the NewAthena Instrument Consortia. The instrument responses used to produce Figs. 3 and 4 are publicly available on the websites of the corresponding missions.

## References

1. Martizzi, D. et al. Baryons in the Cosmic Web of IllustrisTNG—I: gas in knots, filaments, sheets, and voids. *Mon. Not. R. Astron. Soc.* **486**, 3766–3787 (2019).
2. Aartsen, M. G. et al. IceCube-Gen2: the window to the extreme Universe. *J. Phys. G Nucl. Part. Phys.* **48**, 060501 (2021).
3. Adrián-Martínez, S. et al. Letter of intent for KM3NeT 2.0. *J. Phys. G Nucl. Part. Phys.* **43**, 084001 (2016).
4. Amaro-Seoane, P. et al. Astrophysics with the Laser Interferometer Space Antenna. *Living Rev. Relativ.* **26**, 2 (2023).
5. Wilkes, B. J., Tucker, W., Schartel, N. & Santos-Lleo, M. X-ray astronomy comes of age. *Nature* **606**, 261–271 (2022).
6. Magorrian, J. et al. The demography of massive dark objects in galaxy centers. *Astron. J.* **115**, 2285–2305 (1998).
7. King, A. The AGN–starburst connection, galactic superwinds, and $M_{BH}$–$\sigma$. *Astrophys. J.* **635**, L121–L123 (2005).
8. Aird, J. et al. The evolution of the hard X-ray luminosity function of AGN. *Mon. Not. R. Astron. Soc.* **401**, 2531–2551 (2010).
9. Maiolino, R. et al. A small and vigorous black hole in the early Universe. *Nature* **627**, 59–63 (2024).
10. Hopkins, P. F., Hernquist, L., Cox, T. J. & Kereš, D. A cosmological framework for the co-evolution of quasars, supermassive black holes, and elliptical galaxies. I. Galaxy mergers and quasar activity. *Astrophys. J. Suppl. Ser.* **175**, 356–389 (2008).
11. Tombesi, F. et al. Evidence for ultrafast outflows in radio-quiet AGNs. I. Detection and statistical incidence of Fe K-shell absorption lines. *Astron. Astrophys.* **521**, 57 (2010).
12. Fabian, A. C. Observational evidence of active galactic nuclei feedback. *Annu. Rev. Astron. Astrophys.* **50**, 455–489 (2012).
13. Schaye, J. et al. The physics driving the cosmic star formation history. *Mon. Not. R. Astron. Soc.* **402**, 1536–1560 (2010).
14. Hickox, R. C. et al. Black hole variability and the star formation–active galactic nucleus connection: do all star-forming galaxies host an active galactic nucleus? *Astrophys. J.* **782**, 9 (2014).
15. Reynolds, C. S. Observational constraints on black hole spin. *Annu. Rev. Astron. Astrophys.* **59**, 117–154 (2021).
16. Merloni, A. et al. The SRG/eROSITA all-sky survey. First X-ray catalogues and data release of the western Galactic hemisphere. *Astron. Astrophys.* **682**, A34 (2024).
17. Georgakakis, A. et al. X-ray selected AGN in groups at redshifts $z \sim 1$. *Mon. Not. R. Astron. Soc.* **391**, 183–189 (2008).
18. Smail, I., Hogg, D. W., Yan, L. & Cohen, J. G. Deep optical galaxy counts with the Keck Telescope. *Astrophys. J.* **449**, L105 (1995).
19. Vernstrom, T. et al. Deep 3-GHz observations of the Lockman Hole North with the Very Large Array—II. Catalogue and μJy source properties. *Mon. Not. R. Astron. Soc.* **462**, 2934–2949 (2016).
20. Brandt, W. N. & Alexander, D. M. Cosmic X-ray surveys of distant active galaxies. The demographics, physics, and ecology of growing supermassive black holes. *Astron. Astrophys. Rev.* **23**, 93 (2015).
21. Aird, J. et al. The evolution of the X-ray luminosity functions of unabsorbed and absorbed AGNs out to $z \sim 5$. *Mon. Not. R. Astron. Soc.* **451**, 1892–1927 (2015).
22. Mazzolari, G. et al. Heavily obscured AGN detection: a radio versus X-ray challenge. *Astron. Astrophys.* **687**, 120 (2024).
23. Cassano, R. et al. SKA–Athena Synergy White Paper. Preprint at https://arxiv.org/abs/1807.09080 (2018).
24. Padovani, P. et al. ESO–Athena Synergy White Paper. Preprint at https://arxiv.org/abs/1705.06064 (2017).
25. Troja, E. et al. The X-ray counterpart to the gravitational-wave event GW170817. *Nature* **551**, 71–74 (2017).
26. Iacovelli, F., Mancarella, M., Foffa, S. & Maggiore, M. Forecasting the detection capabilities of third-generation gravitational-wave detectors using GWFAST. *Astrophys. J.* **941**, 208 (2022).
27. Nakar, E. & Piran, T. Detectable radio flares following gravitational waves from mergers of binary neutron stars. *Nature* **478**, 82–84 (2011).
28. Piro, L. et al. Athena synergies in the multi-messenger and transient universe. *Exp. Astron.* **54**, 23–117 (2022).
29. Morsink, S. M., Leahy, D. A., Cadeau, C. & Braga, J. The oblate Schwarzschild approximation for light curves of rapidly rotating neutron stars. *Astrophys. J.* **663**, 1244–1251 (2007).
30. Riley, T. E. et al. A NICER view of PSR J0030+0451: millisecond pulsar parameter estimation. *Astrophys. J.* **887**, L21 (2019).
31. Miller, M. C. et al. PSR J0030+0451 mass and radius from NICER data and implications for the properties of neutron star matter. *Astrophys. J.* **887**, L24 (2019).
32. Borgani, S. & Kravtsov, A. Cosmological simulations of galaxy clusters. *Adv. Sci. Lett.* **4**, 204–227 (2011).
33. Tashiro, M. Concept of the X-ray Astronomy Recovery Mission. *Proc. SPIE* **10699**, 1069922 (2018).
34. Ishisaki, Y. et al. Resolve instrument on X-ray Astronomy Recovery Mission (XARM). *J. Low Temp. Phys.* **193**, 991–995 (2018).
35. Brunetti, G. & Jones, T. W. in *Magnetic Fields in Diffuse Media* Vol. 407 (eds Lazarian, A. et al.) 557–598 (Springer, 2015).
36. Predehl, P. et al. Detection of large-scale X-ray bubbles in the Milky Way halo. *Nature* **7837**, 227–231 (2020).
37. Pillepich, A. et al. X-ray bubbles in the circumgalactic medium of TNG50 Milky Way- and M31-like galaxies: signposts of supermassive black hole activity. *Mon. Not. R. Astron. Soc.* **508**, 4667–4695 (2021).
38. Reiprich, T. H. et al. The Abell 3391/95 galaxy cluster system. A 15 Mpc intergalactic medium emission filament, a warm gas bridge, infalling matter clumps, and (re-) accelerated plasma discovered by combining SRG/eROSITA data with ASKAP/EMU and DECam data. *Astron. Astrophys.* **647**, A2 (2021).
39. Truong, N. et al. X-ray metal line emission from the hot circumgalactic medium: probing the effects of supermassive black hole feedback. *Mon. Not. R. Astron. Soc.* **525**, 1976–1997 (2023).
40. Nicastro, F. et al. Observations of the missing baryons in the warm–hot intergalactic medium. *Nature* **558**, 406–409 (2018).

41. Mernier, F. et al. Enrichment of the hot intracluster medium: observations. *Space Sci. Rev.* **214**, 129 (2018).
42. Kobayashi, C., Karakas, A. I. & Lugaro, M. The origin of elements from carbon to uranium. *Astrophys. J.* **900**, 179 (2020).
43. Pass, E. K. et al. Mid-to-late M dwarfs lack Jupiter analogs. *Astron. J.* **166**, 11 (2023).
44. Bryson, S. et al. The occurrence of rocky habitable-zone planets around Solar-like stars from Kepler data. *Astron. J.* **161**, 36 (2021).
45. Ment, K. & Charbonneau, D. The occurrence rate of terrestrial planets orbiting nearby mid-to-late M dwarfs from TESS sectors 1–42. *Astron. J.* **165**, 265 (2023).
46. Nandra, K., et al. The Hot and Energetic Universe: a White Paper presenting the science theme motivating the Athena+ mission. Preprint at https://arxiv.org/abs/1306.2307 (2013).
47. Guainazzi, M. & Tashiro, M. S. The Hot Universe with XRISM and Athena. *Proc. IAU* **342**, 29–36 (2020).
48. Barret, D. et al. The Athena space X-ray observatory and the astrophysics of hot plasma. *Astron. Nachr.* **341**, 224–235 (2020).
49. Reynolds, C. S. et al. Overview of the advanced x-ray imaging satellite (AXIS). *Proc. SPIE* **12678**, 126781E (2023).
50. Bregman, J. et al. Scientific objectives of the Hot Universe Baryon Surveyor (HUBS) mission. *Sci. China Phys. Mech. Astron.* **66**, 299513 (2023).
51. Meidinger, N. et al. Development status of the Wide Field Imager instrument for Athena. *Proc. SPIE* **11444**, 114440T (2020).
52. Barret, D. et al. The Athena X-ray Integral Field Unit: a consolidated design for the system requirement review of the preliminary definition phase. *Exp. Astron.* **55**, 373–426 (2023).
53. Collon, M. J. et al. The development of the mirror for the Athena x-ray mission. *Proc. SPIE* **12181**, 121810U (2022).
54. Bavdaz, M. et al. ATHENA optics technology development. *Proc. SPIE* **12181**, 121810T (2022).
55. Dauser, T. et al. SIXTE: a generic X-ray instrument simulation toolkit. *Astron. Astrophys.* **630**, A66 (2019).
56. Sironi, G. et al. Open-source simulator for ATHENA X-ray telescope optics. *Proc. SPIE* **11822**, 118220I (2021).

## Acknowledgements
We explicitly acknowledge the work of countless scientists at ESA, in the WFI and X-IFU Instrument Consortia and in the International Partners of the Athena Study (JAXA and NASA), as well as in the scientific community coordinated by the Athena Community Office, as a source of continuous inspiration for this manuscript and— more fundamentally—for the Science Redefinition Team (SRDT) contribution to the reformulation of the NewAthena science case. We express our appreciation to scientists and engineers in the aforementioned institutions, who enabled the definition of NewAthena as a technical and financially viable project in the framework of the ESA Science Programme. The SIXTE software package[55], a generic, mission-independent Monte Carlo simulation toolkit for X-ray astronomical instrumentation, has been extensively used to create the figures in this manuscript. Some of the performance simulations for NewAthena have been provided by SIMPOSIuM, an ESA-financed project aimed at developing an open-source silicon pore optics simulation tool[56]. The bottom right panel of Fig. 3 was provided by A. Rau (MPE, Garching). Comments by D. Barret and E. Kuulkers on an earlier version of the manuscript are gratefully acknowledged.

## Author contributions
This manuscript was prepared in a collaborative fashion by all authors, as members of the NewAthena SRDT, appointed by ESA in 2023 to evaluate the science case of the mission. Each of the authors contributed to a specific section or subsection and provided inputs for the generation of the figures. Specific details are the following. F.J.C., M.G. and N.R. coordinated the preparation of the manuscript. They defined the structure of the Perspective, and wrote the manuscript introduction. F.J.C. coordinated the elaboration in 'How do black holes grow and influence galaxy evolution?'. J.A., F.J.C., T.D., D.P. and P.-O.P. contributed to 'How do black holes grow and influence galaxy evolution?'. M.G. coordinated the elaboration of 'Mapping the dynamical assembly of intergalactic plasma in the large-scale structure' and 'Probing the evolution of metal factories in the Universe'. D.E., F.G., G.W.P., T.H.R. and A.S. contributed to 'Mapping the dynamical assembly of intergalactic plasma in the large-scale structure' and 'Probing the evolution of metal factories in the Universe'. N.R. coordinated the elaboration of 'Probing star–planet interaction with accurate X-ray spectroscopy' and 'X-ray emission from neutron stars as a probe of dense-matter physics and multimessenger astrophysics'. N.R., L.C., E.C., H.M., R.O. and E.T. contributed to 'Probing star–planet interaction with accurate X-ray spectroscopy' and 'X-ray emission from neutron stars as a probe of dense-matter physics and multimessenger astrophysics'. M.G. and D.S. contributed to the elaboration of 'NewAthena'. J.A. and F.J.C. prepared Fig. 1. A.S. prepared Fig. 2. M.G. prepared Figs. 3 and 4. M.G. was the main Perspective editor. M.C. was the Chair of the SRDT and coordinated the definition and elaboration of the manuscript. All the authors have revised and provided comments on the manuscript at various stages of its elaboration.

## Competing interests
The authors declare no competing interests.

## Additional information
**Correspondence** should be addressed to Matteo Guainazzi.

**Peer review information** *Nature Astronomy* thanks the anonymous reviewers for their contribution to the peer review of this work.

**Reprints and permissions information** is available at www.nature.com/reprints.

Mike Cruise[1], Matteo Guainazzi ●[2] ✉, James Aird[3], Francisco J. Carrera ●[4], Elisa Costantini[5], Lia Corrales[6], Thomas Dauser[7], Dominique Eckert ●[8], Fabio Gastaldello ●[9], Hironori Matsumoto[10], Rachel Osten ●[11,12], Pierre-Olivier Petrucci ●[13], Delphine Porquet ●[14], Gabriel W. Pratt[15], Nanda Rea ●[16,17], Thomas H. Reiprich ●[18], Aurora Simionescu ●[5], Daniele Spiga[19] & Eleonora Troja ●[20]

[1]School of Physics and Astronomy, University of Birmingham, Birmingham, UK. [2]European Space Agency (ESA), European Space Research and Technology Centre (ESTEC), Noordwijk, The Netherlands. [3]University of Edinburgh Institute for Astronomy, Royal Observatory, Edinburgh, UK. [4]Instituto de Física de Cantabria (CSIC–Universidad de Cantabria), Santander, Spain. [5]SRON Netherlands Institute for Space Research, Leiden, The Netherlands. [6]Astronomy Department, University of Michigan, Ann Arbor, MI, USA. [7]Remeis-Observatory & ECAP, FAU Erlangen-Nürnberg, Bamberg, Germany. [8]Department of Astronomy, University of Geneva, Versoix, Switzerland. [9]INAF/IASF-Milano, Milan, Italy. [10]Department of Earth and Space Science, Graduate School of Science, Osaka University, Toyonaka, Japan. [11]Space Telescope Science Institute, Baltimore, MD, USA. [12]Center for Astrophysical Sciences, Department of Physics and Astronomy, Johns Hopkins University, Baltimore, MD, USA. [13]IPAG, Université Grenoble Alpes, CNRS, Grenoble, France. [14]LAM, Aix Marseille University, CNRS, CNES, Marseille, France. [15]AIM, Université Paris-Saclay, Université Paris Cité, CEA, CNRS, Gif-sur-Yvette, France. [16]Institute of Space Sciences (ICE-CSIC), Barcelona, Spain. [17]Institut d'Estudis Espacials de Catalunya (IEEC), Barcelona, Spain. [18]Argelander Institute for Astronomy, University of Bonn, Bonn, Germany. [19]INAF/Brera Astronomical Observatory, Merate, Italy. [20]Department of Physics, University of Rome Tor Vergata, Rome, Italy. ✉e-mail: Matteo.Guainazzi@esa.int